# A Theory for Diffusionless Structural Phase Transitions in BaTiO$_3$ and PZT Single Crystals

Hui Zhang

School of Materials Science and Engineering, South China University of Technology, Guangzhou 510640, People's Republic of China

Abstract

A phenomenological theory has been proposed for the diffusionless structural phase transitions in crystalline solids here. It has been found that for BaTiO$_3$ single crystal, both the phase transitions and the crystal structures can be predicted with the crystalline anisotropy constants and order parameter dependent strain constants that depend on the temperature. The results have also shown that the morphotropic phase boundaries in PZT crystals arise from the crystalline anisotropy.

## I. Introduction

The mechanism about diffusionless phase transition has received considerable attention due to its great technological importance in steels, ferromagnetic shape memory alloys, and ferroelectric crystals. In the alloys, this phase transition is often referred to as the martensitic phase transition [1]. The investigations on NiMnGa ferromagnetic shape memory alloys have shown that the driving force is closely associated with larger magnetocrystalline anisotropy constants of the martensitic phase [2-5]. Such similar phenomena have also been experimentally observed in $BaTiO_3$ and $PbZrO_3$ -$x$$PbTiO_3$ (PZT) ferroelectric inorganic crystals although the order parameters involved are different [6-11]. For $BaTiO_3$ single crystal, the first-principle calculations by Hu and Cohen have demonstrated that at zero temperature, the rhombohedral phase is energetically stable, but further study is needed for the finite temperature [12]. However, both experimental observations and theoretical results have shown some similarities in these structural phase transitions, which bring to us to a question of whether there is a general theory for these phase transitions. For ferromagnetic shape memory alloys, the decrease of the temperature can induce the cubic → tetragonal phase transition. But for $BaTiO_3$ single crystal, the sequence of phase transitions are more complex and as follows: cubic→ tetragonal→ orthorhombic→ rhombohedral [6-10]. Most of these phase transitions have clearly exhibited the couplings between the order parameter and the crystal lattice. For instance, with the variation in the temperature, the coupling of the order parameter with the crystal lattice leads to the variation in the magnetocrystalline anisotropy constants, and thus the induced phase transition in ferromagnetic shape memory alloys [13-17]. Our recent theoretical results have also shown that the polarization rotation is determined by the crystalline anisotropy in $BaTiO_3$ single crystal [18]. Therefore, with $BaTiO_3$ and PZT as the representatives for these diffusionless phase transitions, we have developed a phenomenological model for investigating the phase transitions. In this model, the phase transitions are energetically induced by the crystalline anisotropy and the relative displacement of the atoms in new phase with respect to that in the parent phase can be predicted. The good agreement between the

theoretical results and experimental observations has shown that our theory is effective in describing the phase transitions.

## II. Phenomenological theory for the phase transitions

### A. The free energy function

On dealing with the phase transitions, several assumptions have been made. (1) At a temperature $T$ below the Curie temperature $T_c$, there exists an exchange coupling keeping the order parameter vector parallel or antiparallel to each other. This exchange coupling, with reference to the exchange coupling between the spins in ferromagnetism, can be regarded as isotropic. (2) The order parameter vector can couple with the crystal lattice, and then an order parameter dependent crystalline anisotropy can be created. The crystal cell can have the cubic symmetry before the coupling of the order parameter with crystal lattice has been created. Then the symmetry of the crystalline anisotropy depends on the crystalline anisotropy constants. When the crystalline anisotropy alone is involved, the order parameter vector will point to the so-called spontaneous directions and the original cubic crystal cell will be deformed. As a result, the symmetry for the resulting crystal cell can be determined by the order parameter dependent strain constants associated with the coupling between the order parameter vector and the crystal lattice. (3) At any temperature $T$ below $T_c$, the saturation values for the order parameter for a single phase are dependent on the temperature and the compositions of the material. There are the multiple saturation values for the order parameters for a certain temperature range. For example, for $BaTiO_3$ single crystal, there is the hysteresis in the saturation polarization vs.

temperature curves for the temperature range of 163~363 *K* [6-10]. When the temperature decreases, the crystal exhibits the saturation polarization of the orthorhombic phase until the orthorhombic→ rhombohedral phase transition temperature. On the contrary, the crystal shows the saturation polarization of the rhombohedral phase until the rhombohedral→ orthorhombic phase transition temperature. (4) The crystal consists of non-interacting single domains. Within each domain, the exchange coupling keeps all the order parameter vector parallel to each other. The oritentations for the order parameter vector for each domain are the spontaneous directions determined by the crystalline anisotropy. (5) When the phase transition occurs, the order parameter vector for single domains in one phase can be accordingly rotated into the new spontaneous order parameter directions in the other phase. Then the crystal cell can be accordingly deformed and this deformation depends on the order parameter dependent strain constants.

We have chosen BaTiO$_3$ and PZT single crystals for investigation due to its complexity in the phase transitions and relatively rich experimental data. Above $T_c$, BaTiO$_3$ single crystal is paraelectric and has the cubic symmetry. Below $T_c$, the phase transitions occur and the sequence are as follows: cubic→ tetragonal→ orthorhombic → rhombohedral. According to Landau phase transition theory, the Gibbs free energy function *G* can be modified as

$$G = a(T - T_c)(P_1^2 + P_2^2 + P_3^2) + b(P_1^4 + P_2^4 + P_3^4) + c(P_1^6 + P_2^6 + P_3^6) \qquad (1)$$

where $a$, $b$, and $c$ are the coefficients for Landau free energy, $T$ is the temperature, and $P_1$, $P_2$, and $P_3$ are the components of the spontaneous saturation polarization along the three principal axes. In Eq. (1), some terms have been intentionally neglected because these terms, in our theory, have been indirectly included in Eq. (1). The crystal's free energy given in Eq. (1) can hold fixed when the operations of cubic symmetry are applied. According to our assumptions, $P_1=P_s\alpha_1$, $P_2=P_s\alpha_2$, and $P_3=P_s\alpha_3$ ($\alpha_1$, $\alpha_2$, and $\alpha_3$ being the direction cosines of the saturation polarization)

$$G = k_0 + k_1\left(\alpha_1^4 + \alpha_2^4 + \alpha_3^4\right) + k_2\left(\alpha_1^6 + \alpha_2^6 + \alpha_3^6\right) \tag{2}$$

where $k_0=a(T-T_c)P_s^2$, $k_1=bP_s^4$, and $k_2=cP_s^6$. With the equations in Appendix, Eq. (2) can be rewritten as

$$G = K_0 + K_1\left(\alpha_1^2\alpha_2^2 + \alpha_2^2\alpha_3^2 + \alpha_3^2\alpha_1^2\right) + K_2\left(\alpha_1^2\alpha_2^2\alpha_3^2\right) \tag{3}$$

where $K_0=k_0+k_1+k_2$, $K_1=-(2k_1+3k_2)$, and $K_2=-3k_2$. The free energy in Eq. (3) also can hold fixed under the cubic symmetry operations, and is equivalent to Eq. (2). With reference to the magnetocrystalline anisotropy constants in ferromagnetism, the coefficients $b$ and $c$ (or $k_1$ and $k_2$, or $K_1$ and $K_2$) can be defined as the order parameter dependent crystalline anisotropy constants here though they are in different equations. And the constant $a$ (or $k_0$, or $K_0$) is isotropic and can be used to characterize the exchange coupling that keeps the order parameter vector parallel to each other. With

both these two expressions, the crystalline anisotropy constants can be calculated from Landau-Devonshire (LD) theory [4].

$$\left. \begin{array}{l} G^{100} = k_0 + k_1 + k_2 \\ G^{110} = k_0 + \dfrac{k_1}{2} + \dfrac{k_2}{4} \\ G^{111} = k_0 + \dfrac{k_1}{3} + \dfrac{k_2}{9} \end{array} \right\} \quad (4)$$

and

$$\left. \begin{array}{l} G^{100} = K_0 \\ G^{110} = K_0 + \dfrac{K_1}{4} \\ G^{111} = K_0 + \dfrac{K_1}{3} + \dfrac{K_2}{27} \end{array} \right\} \quad (5)$$

The values for the saturation polarization along the [100], [110], and [111] axes can be calculated in terms of Eq. (2)

$$\left. \begin{array}{l} \dfrac{dG^{100}}{dP_s} = a(T - T_c) + 2bP_s^2 + 3cP_s^4 = 0 \\ \dfrac{dG^{110}}{dP_s} = 4a(T - T_c) + 4bP_s^2 + 3cP_s^4 = 0 \\ \dfrac{dG^{111}}{dP_s} = 3a(T - T_c) + 2bP_s^2 + cP_s^4 = 0 \end{array} \right\} \quad (6)$$

We have

$$P_s^2 = -\frac{b}{3c} \pm \sqrt{\left(\frac{b}{3c}\right)^2 - \frac{a}{3c}(T-T_c)} \quad , P_s // [100]$$

$$P_s^2 = -\frac{2b}{3c} \pm \sqrt{\left(\frac{2b}{3c}\right)^2 - \frac{4a}{3c}(T-T_c)} \quad , P_s // [110] \quad (7)$$

$$P_s^2 = -\frac{b}{c} \pm \sqrt{\left(\frac{b}{c}\right)^2 - \frac{3a}{c}(T-T_c)} \quad , P_s // [111]$$

When the digital coefficients in Eq. (7) are incorporated into $a$, $b$, and $c$, respectively, the solutions for $P_s^2$ are of the form

$$P_s^2 = -\xi \pm \sqrt{\xi^2 - \zeta(T-T_c)} \quad (8)$$

Table I lists the solutions for $P_s^2$ with various values for the crystalline anisotropy constants $b$ and $c$. It has been shown in Table I that the solutions of saturation polarization are strongly dependent on the crystalline anisotropy constants. At $T<T_c$, and for $a>0$, there is only one non-zero solution for $P_s^2$ for $b>0$ and $c>0$, and $b<0$ and $c>0$, and no non-zero solutions for $b<0$ and $c<0$. More interestingly, there are possibly two non-zero solutions for $b>0$ and $c<0$. Due to the complicated dependence of the crystalline anisotropy constants on the temperature, the solutions listed in Table I are possible. Detailed computation can be found in Appendix.

Table I. The solutions for $P_s^2$ with various values for the crystalline anisotropy constants $b$ and $c$. And $T<T_c$ and $a>0$.

**B. The phase transitions in BaTiO$_3$ single crystal**

According to the theory mentioned above, we have investigated the phase transition sequence in BaTiO$_3$ single crystal. Figure 1(a) shows the experimental data for the dependence of the components of the saturation polarization along the [100] axis on the temperature in BaTiO$_3$ single crystal. Figures (b) and (c) show the calcaluted data for $E$=0 and $E$=500 MV/m along the [100] axis, respectively. It has been indicated in Figs. 1(a) and 1(b) that the numerical results under zero field are in good agreement with the experimental data. Figure 1(d) shows the dependence of the crystalline anisotropy constants $K_1$ and $K_2$ (or $k_1$ and $k_2$) on the temperature. It has been found in Fig. 1 that the phase transitions in BaTiO$_3$ single crystal occur due to the complicated dependence of the crystalline anisotropy constants $K_1$ and $K_2$ (or $k_1$ and $k_2$) on the temperature. When the temperature varies, the changes in the crystalline constants $K_1$ and $K_2$ lead to the changes in the spontaneous polarization directions for BaTiO$_3$ single crystal. For these phase transitions, the numerical results have shown that for the tetragonal phase the polar directions are the six <100> axes, for the orthorhombic phase the twelve <110> axes, and for the rhombohedral phase the eight <111> axes. Table II lists the spontaneous polar directions for the values of $K_1$ and $K_2$. It has been shown in Table II that for different combinations of $K_1$ and $K_2$ values the spontaneous polar directions also are different. For BaTiO$_3$ single crystal, the experimental data have demonstrated that accompanying with the phase transitions there is the variation in the polar directions, and the new polar directions determine the crystal structures of the new phases [6-10]. Therefore, the above numerical results are in good agreement with the experimental observations. When the dependence of the crystalline

anisotropy constants $K_1$ and $K_2$ (or $k_1$ and $k_2$) on the temperature is known, the phase transitions in single crystal can be predicted.

Table II. The spontaneous polar directions for the values of $K_1$ and $K_2$. Easy, Medium, and Hard correspond to the directions with the lowest, lower, and high energy levels calculated according to Eq. (3) [19].

The results also demonstrate the hysteresis when the phase transitions occur. At the critical temperature $T_1$', there is no hysteresis for this tetragonal↔ orthorhombic phase transition but there is the polarization jump. However, at the critical temperatures $T_2$ and $T_2$', there are the orthorhombic↔ rhombohedral phase transitions. For the temperatures between $T_2$ and $T_2$', there coexist two phases (the orthorhombic and rhombohedral phases) at the same temperature. However, these two phases correspond to different free energy levels. When the temperature decreases, BaTiO$_3$ single crystal still remains in the orthorhombic phase, and the polar directions are the <110> axes. At the temperature $T<T_2$, the saturation polarization is abruptly rotated into the <111> axes. The hysteresis occurs due to the energy barrier between the <110> and <111> directions which can disappear for $T\leq T_2$. This is also the case for the rhombohedral→ orthorhombic phase transition.

FIG. 1(Color online). (a) The experimental data for the dependence of the components of the saturation polarization along the [100] axis on the temperature in BaTiO$_3$ single

crystal. (b) and (c) the calcaluted data for $E=0$ and $E=500$ MV/m along the [100] axis, respectively. (d) The dependence of the crystalline anisotropy constants $K_1$ and $K_2$ on the temperature. In (a) and (c), the arrows show the path for the polarization for single domains when the temperature varies. In (b), the red arrows and the text show the critical temperatures for the phase transitions, and Miller's indices indicate the corresponding spontaneous polar directions for single domains. The electric field can be introduced according to the equations in Ref. [18], and the experimental data in (a) is from Ref. [8].

We can calculate the saturation polarizaitons along differents axes, according to the free energy expression based on $k_1$ and $k_2$ (Eq. (2)). However, the saturation polarization is not explicitly included in the free energy expression based on $K_1$ and $K_2$ (Eq. (3)), which can easily show the change in the orientations for the order parameters in crystalline solids. Furthermore, it is experimentally easier to obtain the values for $K_1$ and $K_2$.

**C. The morphotropic phase boundaries in PZT single crystal**

FIG. 2(Color online). Dependence of the crystalline anisotropy constants $K_1$ on the composition $x$ of $PbZrO_3$ -$xPbTiO_3$ single crystals with the temperature of 300 $K$. The inset shows the variation in the values for $K_1$ in the $x$ range of 0.3~0.5.

The morphotropic phase boundaries (MPBs) in PZT crystals can also be understood with our theory. It is well known that for the PZT system Jafe et al. [20] have predicted the PZT crystals with large piezoelectric coefficients when its composition is near the MPB. The experimental observations and theoretical results also have indicated the monoclinic phase of PZT near the MPB [21-23]. According to our theory, the MPB is related to the crystalline anisotropy constants. Figure 2 shows the dependence of the crystalline anisotropy constants $K_1$ on the composition $x$ of PbZrO$_3$-$x$PbTiO$_3$ single crystals with the temperature of 300 $K$. The calcaluted values for $K_2$ are negative for the whole composition range and not shown in Fig. 2. Therefore, the values for $K_1$ can play a key role in determining the phase transition (see Table II). The experimental data have shown that the polar directions for the pure PbZrO$_3$ crystal are the family of <111> axes [24], and for the pure PbTiO$_3$ crystal are the <100> axes [25,26]. It has been indicated in Fig. 2 that their corresponding values for $K_1$ are negative and positive, respectively, in accordance with the predictions listed in Table II. As shown in Fig. 2, the rhombohedral→ tetragonal phase transition occurs at the critical composition of $x_c$=0.38 when the composition of PZT $x$ increases. This rotation of the polar directions from the [111] axes into the [100] axes occurs due to the switch of the signs of $K_1$ values with the composition of PZT. As shown in the inset in Fig. 2, the crystalline anisotropy constant $K_1$ is relatively small (~0.01-0.1 MJ/m$^3$) near the critical composition $x_c$. This means that for PZT crystals with the composition near the MPB the polar directions can be easily disturbed (for instance, by the internal stresses), leading to the significant change in the polar directions. This

can further be understood in combination with single domain rotation model [18]. For single domains, the spontaneous polar directions are determined by the crystalline anisotropy field $E_K$, and for the crystals near MPB this values for $E_K$ can be estimated as $E_K = 2K_1/P_s = 2\times(0.01\text{-}0.1)/0.4 = 0.05\text{-}0.5$ MV/m [18,19]. For the compressive stress strength of 0.01-2MPa, we also can estimate the corresponding stress anisotropy field $E_\sigma = 3\lambda_{100}\sigma/P_s = -3\times0.1\times(0.1\text{-}2)/0.4 = -(0.075\text{-}1.5)$ MV/m [18,19]. We can see that the stress anisotropy field is comparable to (or larger than) the crystalline anisotropy field, and can significantly modify the spontaneous polar directions with the crystal lattice changed. Even though the stress anisotropy is taken into account, the effective anisotropy determined by the combination of both the stress and crystalline anisotropies can still be relatively small when compared to those of PZT crystals far away from the MPB. In this case, the saturation polarization can be rotated into the field directions with small field strength (~1 MV/m). For ferroelectric crystals, this can dramastically enhance the permmitivity $\eta$, and thus the piezoelectric coefficients $d$ ($\propto \eta$) [10].

## III. The induced changes in crystal structures

### A. The atomic displacement in the phase transitions

For the diffusionless phase transition, the atomic displacement in the crystal cell is very important for the formation of new phases. In Section II, we have theoretically demonstrated that the diffusionless transitions are driven by the crystalline anisotropy. At $T<T_c$, the phase transitions occur while the change of the spontaneous order

parameter vector is induced in crystalline solids. The change in the lattice parameters for a crystal cell is dependent on the coupling between the crystal lattice and the order parameter vector. Experimentally, this coupling can be associated with the spontaneous strains that depend on the order parameters. For example, for ferroelectric materials, the coupling can be described by the electrostrictive coefficients, and for ferromagnetic materials by the saturation magnetostriction constants. For crystalline solids, the number of parameters to characterize the coupling may be further reduced to two, i.e., the saturation strains along the [100] and [111] axes for a single crystal such as $\lambda_{100}$ and $\lambda_{111}$. Accompanying with the phase transitions, the relative atomic displacement of new phase with respect to that of the parent phase occurs, making the parent phase with the cubic symmetry deform and transform into the new phase with low symmetry, and vice versa.

FIG. 3(Color online). The geometrical transformation for the atomic coordinates in the phase transition. $r_0(x_0, y_0, z_0)$ represents the vector for the *A* point in the parent phase, and $r(x, y, z)$ represents the vector for the *B* point in the new phase.

The phase with higher crystal symmetry can be regarded as a sphere due to zero values for the order parameters, and the phase with lower crystal symmetry can be seen as an ellipsoid for the anisotropic order parameters. The formation of new phase after the phase transitions is due to the relative atomic displacement of the new phase with respect to the parent phase. Figure 3 shows the geometrical transformation for

the atomic coordinates in the phase transition. It has been shown in Fig. 3 that a sphere can be deformed into an ellipsoid, and point *A* can be accordingly changed to point *B*. For this, a strain translation matrix for the description of the relative atomic displacement of new crystal cell with respect to the old crystal cell can be written as [27]

$$\begin{pmatrix} x \\ y \\ z \end{pmatrix} = \begin{pmatrix} (1+e_{xx}) & \frac{1}{2}e_{xy} & \frac{1}{2}e_{zx} \\ \frac{1}{2}e_{xy} & (1+e_{yy}) & \frac{1}{2}e_{yz} \\ \frac{1}{2}e_{zx} & \frac{1}{2}e_{yz} & (1+e_{zz}) \end{pmatrix} \begin{pmatrix} x_0 \\ y_0 \\ z_0 \end{pmatrix}$$

(9)

where $e_{ij}$ are the strains, and $r_0(x_0, y_0, z_0)$ and $r(x, y, z)$ represent the coordinates of the atoms in crystal cell for the parent phase and the new phase, respectively. Here we simply use the expression given for the coupling between the magnetostrictive strains and order parameters (the magnetization) in ferromagnetism [27,28]

$$\left. \begin{aligned} e_{xx} &= \frac{3}{2}\lambda_{100}\left(\alpha_1^2 - \frac{1}{3}\right) \\ e_{yy} &= \frac{3}{2}\lambda_{100}\left(\alpha_2^2 - \frac{1}{3}\right) \\ e_{zz} &= \frac{3}{2}\lambda_{100}\left(\alpha_3^2 - \frac{1}{3}\right) \\ e_{xy} &= 3\lambda_{111}\alpha_1\alpha_2 \\ e_{zx} &= 3\lambda_{111}\alpha_1\alpha_3 \\ e_{yz} &= 3\lambda_{111}\alpha_2\alpha_3 \end{aligned} \right\}$$

(10)

where $\lambda_{100}$ and $\lambda_{111}$ are the saturation strain coefficients for a single crystal along the [100] and [111] axes. This coupling also can be described by other expressions [18].

Substituting the expression for $e_{ij}$ into Eq. (9), we have the translation matrix for the relative atomic displacement for the crystal cell from the parent phase to the new phase

$$T = \begin{pmatrix} 1+\frac{3}{2}\lambda_{100}\left(\alpha_1^2-\frac{1}{3}\right) & \frac{3}{2}\lambda_{111}\alpha_1\alpha_2 & \frac{3}{2}\lambda_{111}\alpha_1\alpha_3 \\ \frac{3}{2}\lambda_{111}\alpha_1\alpha_2 & 1+\frac{3}{2}\lambda_{100}\left(\alpha_2^2-\frac{1}{3}\right) & \frac{3}{2}\lambda_{111}\alpha_2\alpha_3 \\ \frac{3}{2}\lambda_{111}\alpha_1\alpha_3 & \frac{3}{2}\lambda_{111}\alpha_2\alpha_3 & 1+\frac{3}{2}\lambda_{100}\left(\alpha_3^2-\frac{1}{3}\right) \end{pmatrix} \quad (11)$$

With the translation matrix, we have investigated the phase transition due to the relative atomic displacements in BaTiO$_3$ single crystal when the polarizations are aligned along different spontaneous polar axes.

**B. The phase transition in BaTiO$_3$ single crystal**

Figure 4 shows the variation in the crystal structure and the relative atomic displacement in the phase transition. The sequence of phase transitions are as follows: cubic→ tetragonal→ orthorhombic→ rhombohedral. For tetragonal, orthorhombic, and rhombohedral phases, the spontaneous polarization directions are the <100>, <110>, and <111> axes, respectively. *The paraelectric cubic phase is assumed to be their parent phase. When the strains caused by the temperature are not taken into account, there are the cubic→ tetragonal, cubic→orthorhombic, and cubic→ rhombohedral phase transitions. With the above translation matrix, theoretically, we can calculate the atomic displacements accompanying with the phase transition and*

*demonstrate how the cubic parent phase is transformed into the tetragonal,*

*orthorhombic, and rhombohedral phases.*

FIG. 4(Color online). The variation in the crystal structure and the relative atomic displacement in the phase transition.

### 1. Cubic→ Tetragonal phase transition

For the tetragnoal phase, the spontaneous polar directions are the six <100> axes. For the [100] axis, the direction cosines ($\alpha_1$, $\alpha_2$, $\alpha_3$) are (1, 0, 0). Substituting into Eq. (11), the translation matrix can be calculated as

$$T_{C \to T} = \begin{pmatrix} 1+\lambda_{100} & 0 & 0 \\ 0 & 1-\frac{1}{2}\lambda_{100} & 0 \\ 0 & 0 & 1-\frac{1}{2}\lambda_{100} \end{pmatrix} \quad (12)$$

We can also calculate the atomic coordinates for the parent and new phases following Eqs. (9) and (12). Table III lists the variation in the atomic coordinates in BaTiO$_3$ single crystal in the phase transition.

### 2. Cubic→ Rhombohedral phase transitions

For the rhombohedral phase, the spontaneous polar directions are the eight <111> axes. For the [111] axis, the direction cosines ($\alpha_1$, $\alpha_2$, $\alpha_3$) are (1, 1, 1)/√3. The

translation matrix is

$$T_{C \to R} = \begin{pmatrix} 1 & \frac{1}{2}\lambda_{111} & \frac{1}{2}\lambda_{111} \\ \frac{1}{2}\lambda_{111} & 1 & \frac{1}{2}\lambda_{111} \\ \frac{1}{2}\lambda_{111} & \frac{1}{2}\lambda_{111} & 1 \end{pmatrix} \tag{13}$$

### 3. Cubic→ Orthorhombic phase transitions

For the orthorhombic phase, the spontaneous polar directions are the twelve <110> axes. For the [110] axis, the direction cosines ($\alpha_1$, $\alpha_2$, $\alpha_3$) are (1, 1, 0)/√2. The translation matrix is.

$$T_{C \to O} = \begin{pmatrix} 1+\frac{1}{4}\lambda_{100} & \frac{3}{4}\lambda_{111} & 0 \\ \frac{3}{4}\lambda_{111} & 1+\frac{1}{4}\lambda_{100} & 0 \\ 0 & 0 & 1-\frac{1}{2}\lambda_{100} \end{pmatrix} \tag{14}$$

Table III. The variation in the atomic coordinates in BaTiO$_3$ single crystal in the phase transition.

It has been found from the transformation matrix that the relative atomic displacement in crystalline solids is strongly dependent on the strain constants $\lambda_{100}$ and $\lambda_{111}$. For BaTiO$_3$ single crystal, the atomic relative displacement for different phases is listed in Table III. It has been seen in Table III that for tetragonal phase the atomic coordinates

are only dependent on $\lambda_{100}$. For rhombohedral phase the atomic coordinates are only dependent on $\lambda_{111}$. For orthorhombic phase the atomic coordinates are closely associated with both $\lambda_{100}$ and $\lambda_{111}$ whose values can largely influence the crystal structure of new phase. For tetragonal BaTiO$_3$ single crystal, $\lambda_{100}>0$ and $\lambda_{111}>0$ [18,19]. For the cubic→ tetragonal phase transition, the original cubic cell can elongate in the [100] axis and contract in the two [010] and [001] axes, thus leading to the tetragonal cell. Assuming that the lattice constants for the original cubic cell are $a_0$, the lattice constants for the resulting tetragonal cell are $a_1$ and $c_1$, and, obviously, $a_1<c_1$. If $\lambda_{100}<0$, $a_1>c_1$. For orthorhombic phase, the dependence of the crystal structure on $\lambda_{100}$ and $\lambda_{111}$ are comoplex. As shown in Table III, the original cubic cell can contract in one direction and elongate in the other two directions, leading to the orthorhombic cell ($a>c$). Fo rhombohedral phase, the polar directions are the family of <111> axes, suggesting that the cubic cell can elongate or contract in the three directions, and the interaxial angles of the lattice constants can decrease or increase. The results are in good agreement with the experimental data [6,7]. It also has been implied that the new phases with lower symmetries can be obtained when the new spontaneous polar directions are of lower symmetry.

## IV. Conclusions

Our results have indicated that for BaTiO$_3$ single crystal, the phase transitions and the crystal structures can be predicted with the crystalline anisotropy constants and order parameter dependent strain constants that depend on the temperature. The results have

also shown that the morphotropic phase boundaries in PZT crystal arise from the crystalline anisotropy.

**Acknowledgements**

This work is supported by the Fundamental Research Funds for the Central Universities of China (Grant No. 2012ZM0010) and Project 11204087 supported by National Natural Science Foundation of China.

**Appendix**

**A. Translation equations**

$$\left(\alpha_1^6 + \alpha_2^6 + \alpha_3^6\right) = 1 - 3\left(\alpha_1^2\alpha_2^2 + \alpha_2^2\alpha_3^2 + \alpha_3^2\alpha_1^2\right) - 3\alpha_1^2\alpha_2^2\alpha_3^2 \tag{A.1}$$

$$\left(\alpha_1^4 + \alpha_2^4 + \alpha_3^4\right) = 1 - 2\left(\alpha_1^2\alpha_2^2 + \alpha_2^2\alpha_3^2 + \alpha_3^2\alpha_1^2\right) \tag{A.2}$$

**B. Calculating the saturation polarization $P_s$ and the crystalline anisotropy constants $K_1$ and $K_2$**

Landau free energy $G_{LD}$ can be expressed as [25,26]

$$\begin{aligned} G_{LD} &= a_1\left(P_1^2 + P_2^2 + P_3^2\right) + a_{11}\left(P_1^4 + P_2^4 + P_3^4\right) + a_{12}\left(P_1^2P_2^2 + P_2^2P_3^2 + P_3^2P_1^2\right) + \\ &\quad a_{111}\left(P_1^6 + P_2^6 + P_3^6\right) + a_{112}\left[P_1^4\left(P_2^2 + P_3^2\right) + P_2^4\left(P_3^2 + P_1^2\right) + P_3^4\left(P_1^2 + P_2^2\right)\right] + a_{123}P_1^2P_2^2P_3^2 \end{aligned} \tag{A.3}$$

where $a_1$, $a_{11}$, $a_{12}$, $a_{111}$, $a_{112}$, and $a_{123}$ are Landau free energy coefficients, and $P_1$, $P_2$, and $P_3$ are the components of the spontaneous saturation polarization $P_s$ ($P_s=(P_1+P_2+P_3)^{1/2}$).

To calculate the saturation polarization $P_s$ and the crystalline anisotropy constants $K_1$

and $K_2$, we can first obtain the expressions for Landau free energies $G_{LD}^{100}$, $G_{LD}^{110}$, and $G_{LD}^{111}$ along the [100], [110], and [111] directions, respectively. $G_{LD}^{100}$, $G_{LD}^{110}$, and $G_{LD}^{111}$ are also the functions of the saturation polarization $P_s$. Then according to $dG/dP_s = 0$ and $d^2G/dP_s^2 > 0$, we can obtain the values for $P_s$ at the temperature $T<T_c$. Substituting the corresponding values of $P_s$ into $G_{LD}^{100}$, $G_{LD}^{110}$, and $G_{LD}^{111}$, we can have their numerical values. The crystalline anisotropy constants can be calculated from [19]:

$$\left.\begin{aligned} G^{100} &= G_{LD}^{100} = K_0 \\ G^{110} &= G_{LD}^{110} = K_0 + K_1/4 \\ G^{111} &= G_{LD}^{111} = K_0 + K_1/3 + K_2/27 \end{aligned}\right\} \quad (A.4)$$

where Landau free energy coefficients can be found in Ref. 29 for $BaTiO_3$ and Refs. 25 and 26 for $PbZrO_3$-$x$$PbTiO_3$.

**C. Calculating the electrostrictive coefficients $\lambda_{100}$ and $\lambda_{111}$ [18]**

$$\lambda_{100} = 2/3(Q_{11}-Q_{12}) \times P_s^2 \quad (A.5)$$

$$\lambda_{111} = 1/3 Q_{44} \times P_s^2 \quad (A.6)$$

FIG. 1(Color online). (a) The experimental data for the dependence of the components of the saturation polarization along the [100] axis on the temperature in BaTiO$_3$ single crystal. (b) and (c) the calcaluted data for $E=0$ and $E=500$ MV/m along the [100] axis, respectively. (d) The dependence of the crystalline anisotropy constants $K_1$ and $K_2$ on the temperature. In (a) and (c), the arrows show the path for the polarization for single domains when the temperature varies. In (b), the red arrows and the text show the critical temperatures for the phase transitions, and Miller's indices indicate the corresponding spontaneous polar directions for single domains. The electric field can be introduced according to the equations in Ref. [18], and the experimental data in (a) is from Ref. [8].

FIG. 2(Color online). Dependence of the crystalline anisotropy constants $K_1$ on the composition $x$ of PbZrO$_3$ -$x$PbTiO$_3$ single crystals with the temperature of 300 $K$. The inset shows the variation in the values for $K_1$ in the $x$ range of 0.3~0.5.

FIG. 3(Color online). The geometrical transformation for the atomic coordinates in the phase transition. $r_0(x_0, y_0, z_0)$ represents the vector for the $A$ point in the parent phase, and $r(x, y, z)$ represents the vector for the $B$ point in the new phase.

FIG. 4(Color online). The variation in the crystal structure and the relative atomic displacement in the phase transition.

Table I. The solutions for $P_s^2$ with various values for the crystalline anisotropy constants $b$ and $c$. And $T<T_c$ and $a>0$.

| $b$ | $c$ | $\xi=b/c$ | $\zeta=a/c$ | $\xi^2-\zeta(T-T_c)$ | Number | $P_s^2$ |
|---|---|---|---|---|---|---|
| >0 | >0 | >0 | >0 | >$\xi$ | 1 | $-\xi+[\xi^2-\zeta(T-T_c)]^{1/2}$ |
| >0 | <0 | <0 | <0 | <$\xi$ | 2 | $-\xi\pm[\xi^2-\zeta(T-T_c)]^{1/2}$ |
| <0 | >0 | <0 | >0 | >$\xi$ | 1 | $-\xi+[\xi^2-\zeta(T-T_c)]^{1/2}$ |
| <0 | <0 | >0 | <0 | <$\xi$ | 0 | |

Table II. The spontaneous polar directions for the values of $K_1$ and $K_2$. Easy, Medium, and Hard correspond to the directions with the lowest, lower, and high energy levels calculated according to Eq. (3) [19].

| $K_1$ | + | + | + | - | - | - |
|---|---|---|---|---|---|---|
| $K_2$ | +∞ to -9$K_1$/4 | -9$K_1$/4 to -9$K_1$ | -9$K_1$ to -∞ | -∞ to 9\|$K_1$\|/4 | 9\|$K_1$\|/4 to 9\|$K_1$\| | 9\|$K_1$\| to +∞ |
| Easy | <100> | <100> | <111> | <111> | <110> | <110> |
| Medium | <110> | <111> | <100> | <110> | <111> | <100> |
| Hard | <111> | <110> | <110> | <100> | <100> | <111> |

Table III. The variation in the atomic coordinates in $BaTiO_3$ single crystal in the phase transition.

| phase | cubic | | | tetragonal | | | rhombohedral | | | orthorhombic | | |
|---|---|---|---|---|---|---|---|---|---|---|---|---|
| | $x_0$ | $y_0$ | $z_0$ | $x-x_0$ | $y-y_0$ | $z-z_0$ | $x-x_0$ | $y-y_0$ | $z-z_0$ | $x-x_0$ | $y-y_0$ | $z-z_0$ |
| $Ba^{2+}$ | 0 | 0 | 0 | 0 | 0 | 0 | 0 | 0 | 0 | 0 | 0 | 0 |
| $Ti^{4+}$ | 1/2 | 1/2 | 1/2 | $a_1$ | $-a_1/2$ | $-a_1/2$ | $a_2$ | $a_2$ | $a_2$ | $(a_3+b_3)/2$ | $(a_3+b_3)/2$ | $-a_3$ |
| $O^{2-}$ | 1/2 | 1/2 | 0 | $a_1$ | $-a_1/2$ | 0 | $a_2/2$ | $a_2/2$ | $a_2$ | $(a_3+b_3)/2$ | $(a_3+b_3)/2$ | 0 |
| | 1/2 | 0 | 1/2 | $a_1$ | 0 | $-a_1/2$ | $a_2/2$ | $a_2$ | $a_2/2$ | $a_3/2$ | $b_3/2$ | $-a_3$ |
| | 0 | 1/2 | 1/2 | 0 | $-a_1/2$ | $-a_1/2$ | $a_2$ | $a_2/2$ | $a_2/2$ | $b_3/2$ | $a_3/2$ | $-a_3$ |

$a_1=\lambda_{100}/2$, $a_2=\lambda_{111}/2$, $a_3=\lambda_{100}/4$, $b_3=3\lambda_{111}/4$

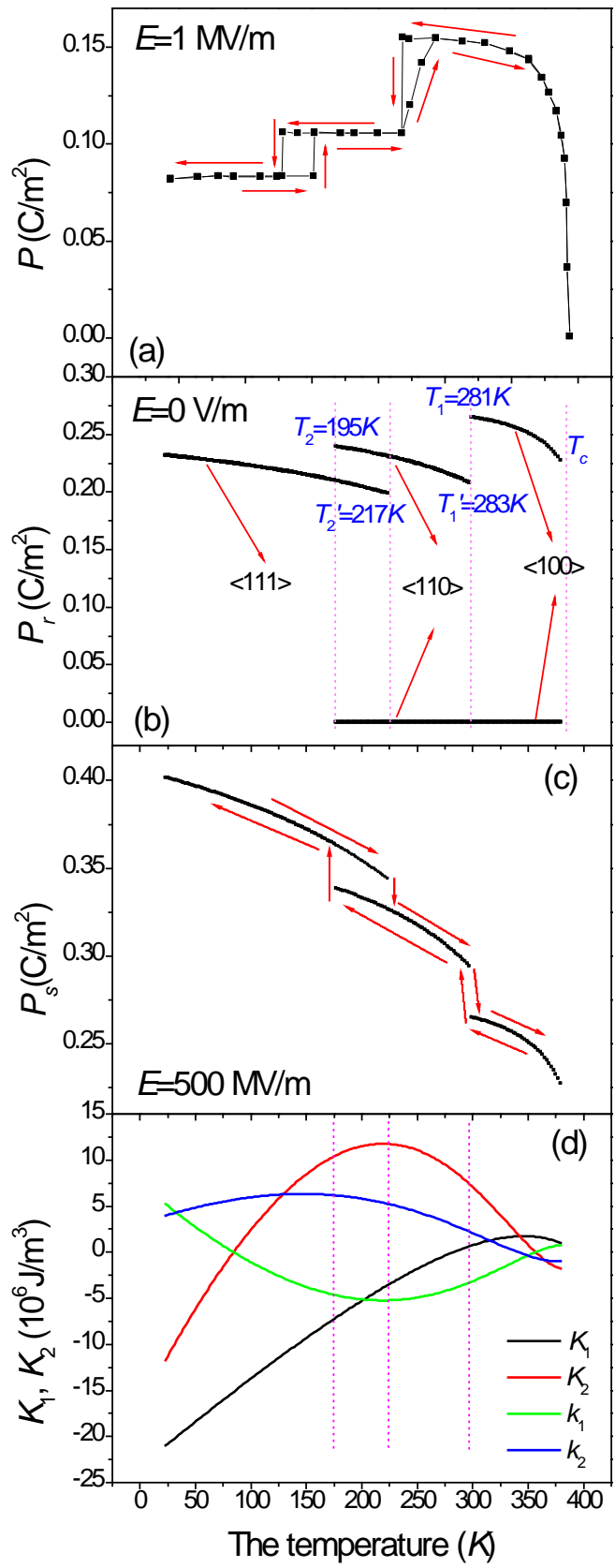

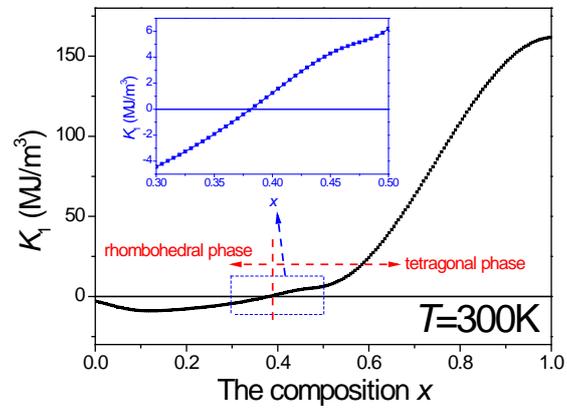

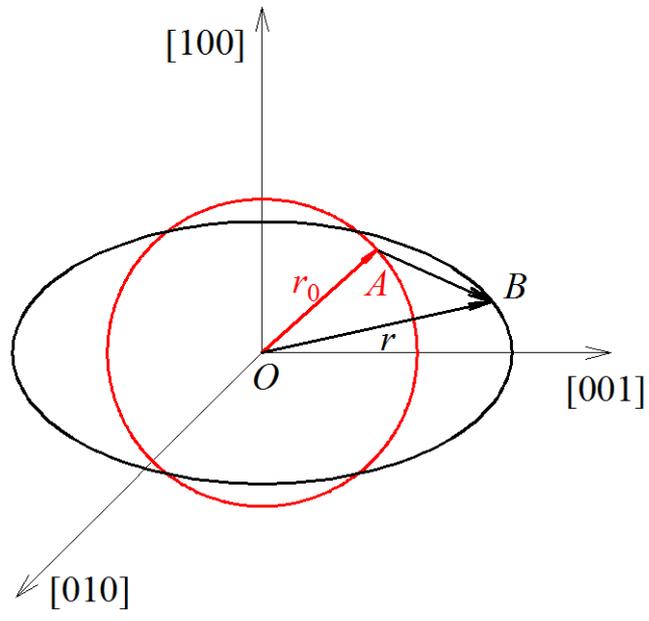

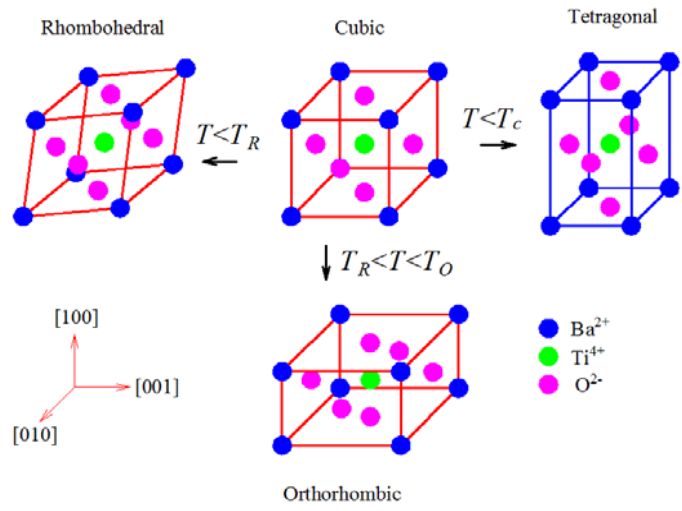